\documentclass[twocolumn,showpacs,amsmath,amssymb]{revtex4}
\usepackage{graphicx}% Include figure files
\usepackage{dcolumn}% Align table columns on decimal point
\usepackage{bm}% bold math

\begin{document}

\preprint{APS/123-QED}

\title{Phase competition %and a temperature independent behavior
in the double-exchange model on the frustrated pyrochlore lattice
}% Force line breaks with \\

\author{
Yukitoshi Motome$^1$ and
Nobuo Furukawa$^{2,3}$
}
\affiliation{%
$^1$Department of Applied Physics, University of Tokyo, Tokyo 113-8656, Japan
\\
$^2$Department of Physics and Mathematics, Aoyama Gakuin University, Kanagawa 229-8558, Japan
\\
$^3$Multiferroics Project, ERATO, Japan Science and Technology Agency (JST)
}%

%\affiliation{

\date{\today}% It is always \today, today,

\begin{abstract}
Competition between the ferromagnetic double-exchange interaction and 
the super-exchange antiferromagnetic interaction is theoretically studied 
in the presence of geometrical frustration.
As increasing the super-exchange interaction, 
the ferromagnetic metal becomes unstable, and 
is taken over by a cooperative paramagnetic metal, 
in sharp contrast with a discontinuous transition to 
the antiferromagnetic insulator in the absence of frustration.
In the critical region, 
the system exhibits a peculiar temperature-independent behavior 
with highly incoherent transport, 
suggesting a large residual entropy at low temperatures. 
We discuss the relevance of the results to 
the pressure-induced behaviors in Mo pyrochlore oxides 
[S. Iguchi {\it et al.}, Phys. Rev. Lett. {\bf 102}, 136407 (2009)].
\end{abstract}

\maketitle

%%%%% Introduction

Interplay between spin and charge degrees of freedom is a central issue 
in the field of strongly-correlated electron systems. 
A main goal is to clarify how magnetic and charge fluctuations are related with each other and 
how various phase transitions 
such as magnetic ordering and metal-insulator transition 
emerges from the interplay. 

One of the fundamental models describing the interplay explicitly 
is the double-exchange (DE) model 
\cite{Zener1951}, 
which has been extensively studied for understanding 
the physics of colossal magnetoresistance (CMR) in perovskite manganese oxides
\cite{Tokura2000}. 
It has been clarified that 
the model exhibits a first-order phase transition with a bicritical behavior 
between a ferromagnetic metal (FM) 
and an antiferromagnetic (AF) insulator 
\cite{Yunoki1998b,Sen2006}. 
The FM is stabilized by the DE interaction
which is an effective ferromagnetic interaction originating from 
the strong coupling between itinerant electrons and localized moments, 
while the AF insulator is stabilized by the super-exchange (SE) 
AF interaction between localized moments. 
It was also discussed that the FM state competes with a charge- and/or orbital-ordered insulator
in extended DE models including the electron-lattice coupling and the orbital degeneracy
\cite{Yunoki2000,Motome2003}. 

In general, magnetism is strongly influenced 
by the geometry of underlying lattice structure. 
In the absence of geometrical frustration, 
AF interaction leads to a simple N\'eel-type AF ordering, 
whereas it becomes unstable against the frustration, resulting in nontrivial phenomena 
such as a complicated ordering, a glassy state, and a spin-liquid state
\cite{Diep2005}.
The frustrated magnetism in localized spin systems is an interesting long-standing issue, 
but it will be more intriguing 
to clarify the effect of geometrical frustration in spin-charge coupled systems. 
The interplay between the frustrated magnetism and the electronic transport 
is expected to yield characteristic spin-charge entangled phenomena. 

\begin{figure}[t]
 \centerline{\includegraphics[width=7truecm]{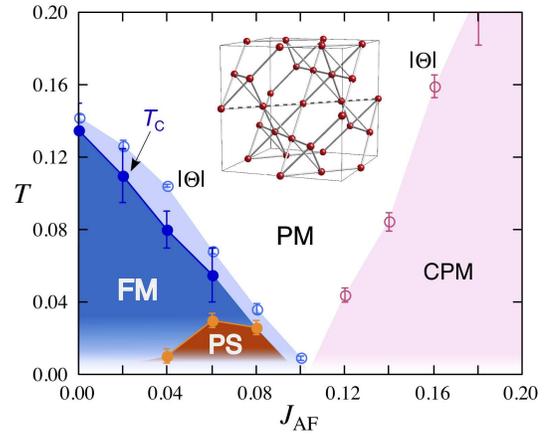}}
\caption{
(Color online)
Phase diagram of the pyrochlore double-exchange model at quarter filling. 
FM, PS, PM, and CPM represent ferromagnetic metal, phase separation, 
paramagnetic metal, and cooperative paramagnetic metal, respectively. 
Closed symbols represent the phase transition boundaries, 
while open ones indicate the magnitude of Curie-Weiss temperature. 
The lines are guides for the eye.
The inset shows the cubic unit cell of the pyrochlore lattice.
The dotted bond shows an example of the 1D chain for the calculations 
of the spin correlation in Fig.~\ref{fig:spincorr}.
}
\label{fig:phase}
\end{figure}

There are many compounds in which the spin and charge degrees of freedom 
couple with each other under frustration.
Among them, Mo pyrochlore oxides {\it R}$_2$Mo$_2$O$_7$ are interesting 
because they exhibit phase competition between FM and spin-glass insulator (SGI)
by changing the rare earth $R$ 
\cite{Ali1989,Katsufuji2000}. 
In these compounds, Mo cations constitute the severely-frustrated pyrochlore lattice, 
which consists of the 3D network of corner-sharing tetrahedra, 
as shown in the inset of Fig.~\ref{fig:phase}. 
It was pointed out by the first-principle band calculation that 
FM is stabilized by the DE interaction, 
while the transition to SGI is triggered by the Coulomb repulsion between Mo $4d$ electrons
and the resulting Mott localization 
\cite{Solovyev2003}. 

Recently, 
several new features were revealed in the Mo pyrochlore oxides under external pressure 
\cite{Mirebeau2006,Iguchi2009}.
The pressure induces a transition from FM to 
a peculiar paramagnetic metallic (PM) state, 
in which the resistivity is highly incoherent and almost temperature($T$) independent. 
In addition, a spin-glass metallic (SGM) phase is found in between. 
There is no metal-insulator transition, 
which is clearly distinguished from the situation in the $R$-site substitution at ambient pressure.
It is highly desired to elucidate relevant interactions 
to the pressure-induced behaviors. 

Motivated by both the theoretical and the experimental problems, in the present study, 
we investigate the phase competition in the DE model defined 
on the 3D frustrated pyrochlore lattice. 
We employ the unbiased Monte Carlo simulation 
and illuminate distinctive aspects of the frustrated spin-charge coupled system.

%%%%% Model

The Hamiltonian of the DE model is given by 
\begin{equation}
{\cal H} = - t \sum_{\langle ij \rangle \sigma}
( c_{i \sigma}^\dagger c_{j \sigma} + {\rm h.c.} )
- J_{\rm H} \sum_i {\mathbf s}_i \cdot {\mathbf S}_i 
+ J_{\rm AF} \sum_{\langle ij \rangle} {\mathbf S}_i \cdot {\mathbf S}_j,
\label{eq:H}
\end{equation}
where $t$ is the transfer integral for nearest-neighbor (n.n.) sites $\langle ij \rangle$ 
on the pyrochlore lattice in the inset of Fig.~\ref{fig:phase}, 
$J_{\rm H}$ represents the Hund's-rule coupling between 
the itinerant electron spin ${\mathbf s}_i$ and the localized spin ${\mathbf S}_i$, 
and $J_{\rm AF}$ is the SE interaction between the neighboring localized spins. 
Some possible symmetry-broken states were examined for this model 
by mean-field approximation 
\cite{Ikoma2003}. 
For simplicity, hereafter, we consider the limit of $J_{\rm H} \to \infty$ and 
treat the localized spins ${\mathbf S}_i$ as classical vectors 
with a normalized length $|{\mathbf S}_i| = 1$ 
\cite{Anderson1955}.
We set an energy unit as $t=1$.
In the following, we investigate the properties of the model (\ref{eq:H}) 
at quarter filling ($0.5$ electron per site on average)
\cite{note}. 

We employ the Monte Carlo simulation to study the thermodynamic properties 
of the model (\ref{eq:H}). 
The algorithm is a standard one in which 
configurations of classical localized spins are sampled by Monte Carlo procedure; 
the Monte Carlo weight is calculated by the exact diagonalization of 
the fermion Hamiltonian matrix for a given spin configuration
\cite{Yunoki1998a}. 
The system sizes $N_{\rm s}$ are 16, 32, 64, and 128 sites 
which correspond to $1 \! \times 1 \! \times \! 1$, 
$1 \! \times 1 \! \times \! 2$, $2 \! \times 2 \! \times \! 1$, and 
$2 \! \times 2 \! \times \! 2$ cubic unit cells. 
We take an average over the twisted boundary conditions
\cite{Poilblanc1991,Gros1992}, 
which enable systematic analysis within the small size clusters down to $T \simeq 0.01$.
The details and the efficiency of this technique are found in Ref.
\cite{Motome2010}.

%%%%% Results and Discussions

\begin{figure}[t]
 \centerline{\includegraphics[width=8truecm]{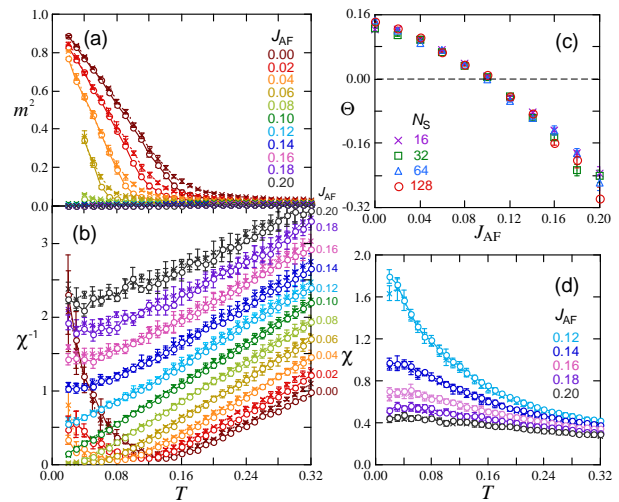}}
\caption{
(Color online)
Temperature dependences of (a) the squared magnetic moment 
and (b) the inverse of the uniform magnetic susceptibility. 
(c) Curie-Weiss temperature estimated from the fitting of $\chi$ 
for various system sizes $N_{\rm s}$.
(d) Temperature dependence of the uniform magnetic susceptibility for large $J_{\rm AF}$.
In (a), (b), and (d), circles and crosses are the data for $N_{\rm s} = 128$ and $64$, respectively.
}
\label{fig:magnetic}
\end{figure}

Figure~\ref{fig:phase} shows the phase diagram of the model (\ref{eq:H}) at quarter filling. 
The ferromagnetic transition temperature $T_{\rm C}$ decreases gradually 
as increasing $J_{\rm AF}$, and approaches zero at $J_{\rm AF} \sim 0.1$. 
Here, $T_{\rm C}$ is estimated by a systematic analysis of the Binder parameter
\cite{Motome2010}. 
After the FM state fades out, we do not find any clear sign of phase transition 
as will be discussed below. 
(We will mention about $|\Theta|$ and the CPM state later.) 
These behaviors are contrastive to the results in the absence of frustration, 
where a bicritical behavior appears with a first-order transition to an AF insulating state
\cite{Yunoki1998b,Sen2006}. 
In the present model, the frustration of the pyrochlore structure is severe enough 
to suppress the AF ordering and replace it with a paramagnetic state. 

Instead of the bicritical behavior in the unfrustrated cases,
an electronic phase separation (PS) takes place at low $T$ as shown in the phase diagram. 
In this region, the system is phase-separated into 
FM at a lower density and PM at a higher density, 
that is, quarter filling is no longer stable:
PS is identified by a jump of the electron density as a function of the chemical potential. 
This metal-to-metal phase separation is a characteristic feature due to the frustration 
which suppresses AF ordering, 
however, we focus on the competition between FM and PM at quarter filling 
in the following, 
except for a brief comment related to experiments later.  
The details of PS will be discussed elsewhere.

In Figs.~\ref{fig:magnetic}(a) and \ref{fig:magnetic}(b), 
we show $T$ dependences of the squared moment 
$m^2 =  \langle (\sum_i {\mathbf S}_i / N_{\rm s})^2 \rangle$
and the inverse of the uniform magnetic susceptibility, $\chi^{-1}$. 
In the small $J_{\rm AF}$ region, 
the ferromagnetic moment grows rapidly as decreasing $T$, and
at the same time $\chi$ shows a peak whose height grows with the system size
\cite{Motome2010}: 
These signal the ferromagnetic transitions. 
As $J_{\rm AF}$ increases, however, the growth of $m^2$ is suppressed gradually, 
corresponding to the collapse of FM in Fig.~\ref{fig:phase}. 
In the larger-$J_{\rm AF}$ region, $m^2$ becomes zero and $\chi$ is suppressed. 

As shown in Fig.~\ref{fig:magnetic}(b), $\chi$ exhibits a Curie-Weiss like behavior at high $T$; 
$\chi^{-1} \propto T-\Theta$, where $\Theta$ is the Curie-Weiss temperature. 
The estimate of $\Theta$, obtained by fitting the high-$T$ part, is plotted 
in Fig.~\ref{fig:magnetic}(c). 
$\Theta$ decreases from positive to negative continuously as increasing $J_{\rm AF}$, 
and crosses zero at $J_{\rm AF} \sim 0.1$. 
This indicates that the effective magnetic interaction changes from 
ferromagnetic to antiferromagnetic continuously, and 
vanishes near the critical region at $J_{\rm AF} \sim 0.1$
where $T_{\rm C}$ goes to zero. 
This suggests a cancellation between the DE ferromagnetic interaction and the SE AF interaction. 

\begin{figure}[t]
 \centerline{\includegraphics[width=8truecm]{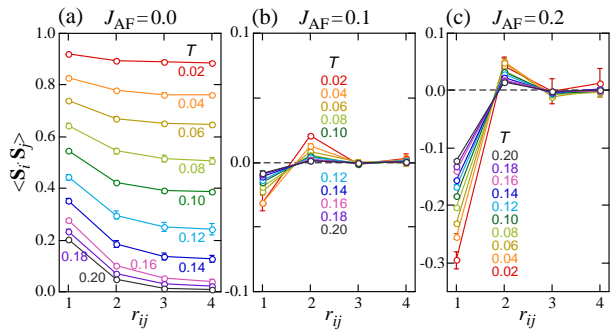}}
\caption{
(Color online)
Spin correlation $\langle {\mathbf S}_i \cdot {\mathbf S}_j \rangle$ 
along the 1D chains in the pyrochlore structure (see the inset of Fig.~\ref{fig:phase})
as a function of distance (measured in unit of the nearest-neighbor bond length)
for (a) $J_{\rm AF} = 0.0$, (b) 0.1, and (c) 0.2.
The data are for $N_{\rm s} = 128$.
}
\label{fig:spincorr}
\end{figure}

Such cancellation is also observed in the spin correlation. 
Figure~\ref{fig:spincorr} shows the distance-dependence of the spin correlation 
$\langle {\mathbf S}_i \cdot {\mathbf S}_j \rangle$ 
for three typical values of $J_{\rm AF} = 0.0$, $0.1$, and $0.2$, corresponding to 
FM, critical, and PM regions, respectively.
In the FM region [Fig.~\ref{fig:spincorr}(a)], 
the spin correlation grows rapidly as decreasing $T$, and 
converges to a positive value for large distance. 
In the critical region at $J_{\rm AF} \sim 0.1$, however, 
the spin correlation hardly develops as shown in Fig.~\ref{fig:spincorr}(b). 
This indicates that the spins remain disordered 
due to the cancellation between the DE ferromagnetic interaction and the SE AF interaction. 

In the PM region with large $J_{\rm AF}$, 
AF correlation develops only for short range at low $T$ [Fig.~\ref{fig:spincorr}(c)]. 
This is remarkable since $\Theta$ increases its magnitude with $J_{\rm AF}$, 
as shown in Fig.~\ref{fig:magnetic}(c), 
indicating a large effective AF interaction $J_{\rm eff} \sim \Theta/8$.
The suppressed AF correlation as well as no clear sign of phase transition 
is due to the strong frustration inherent to the pyrochlore lattice structure. 
This type of frustrated paramagnetic state below $T \sim |\Theta|$ is often called 
a cooperative paramagnet or a classical spin liquid
\cite{Villain1979,Moessner1998}. 
We call the region as cooperative paramagnetic metal (CPM)
and indicate it in Fig.~\ref{fig:phase} by plotting $|\Theta |$.
Note that the susceptibility in the CPM region exhibits a characteristic behavior at low $T$; 
it deviates from the high-$T$ Curie-Weiss behavior and 
shows a broad peak at $T \sim J_{\rm eff}$, as plotted in Fig.~\ref{fig:magnetic}(d). 

\begin{figure}[t]
 \centerline{\includegraphics[width=8truecm]{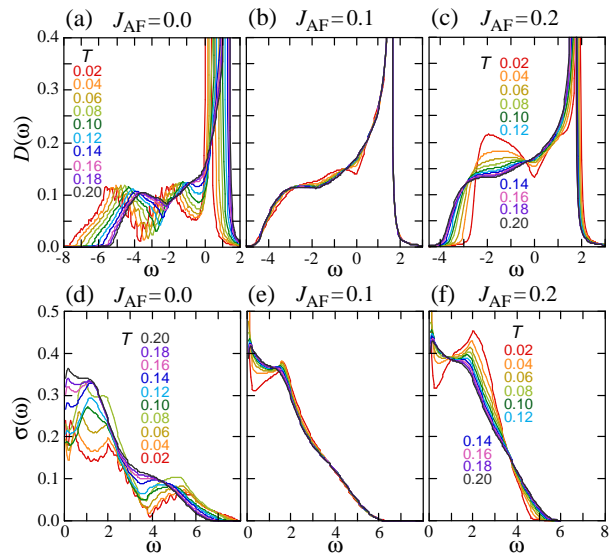}}
\caption{
(Color online)
(a)-(c) Density of states and 
(d)-(f) optical conductivity.
The data are for $N_{\rm s} = 128$.
}
\label{fig:electronic}
\end{figure}

These drastic changes in the magnetic sector affect the electronic state 
through the strong spin-charge coupling in the model (\ref{eq:H}). 
In Fig.~\ref{fig:electronic}, we present the density of states (DOS) $D(\omega)$ 
and the optical conductivity $\sigma(\omega)$ 
for the same set of $J_{\rm AF}$ in Fig.~\ref{fig:spincorr}. 
DOS is plotted by setting the chemical potential to be zero, 
and $\sigma(\omega)$, a diagonal element along the cubic axis, is calculated by 
the standard Kubo formula. 
In the FM region [Figs.~\ref{fig:electronic}(a) and \ref{fig:electronic}(d)], 
both $D(\omega)$ and $\sigma(\omega)$ show a strong $T$ dependence, 
corresponding to the development of ferromagnetism. 
Note that the low-$T$ results are similar to those for the noninteracting model 
(two dispersive bands for $-8 \le \omega \le -4$ and $-4 \le \omega \le 0$, and 
two degenerate flat bands at $\omega=0$)
because of less spin scattering from almost fully-aligned moments. 
(The spiky structure seen in the dispersive bands is due to the finite-size effect.) 
On the other hand, in the large-$J_{\rm AF}$ PM region 
[Figs.~\ref{fig:electronic}(c) and \ref{fig:electronic}(f)], 
$T$ dependence is suppressed but still remains at low $T$, 
reflecting the growth of short-range AF correlations in Fig.~\ref{fig:spincorr}(c)
\cite{note2}. 
In the critical region, neither $D(\omega)$ nor $\sigma(\omega)$ exhibits 
any significant $T$ dependence 
as exemplified in Figs.~\ref{fig:electronic}(b) and \ref{fig:electronic}(e), 
corresponding to the suppression of spin correlation down to the lowest $T$ 
in Fig.~\ref{fig:spincorr}(b). 
Here the electrons are scattered by almost uncorrelated moments, 
resulting in the highly-incoherent conductivity 
shown in Fig.~\ref{fig:electronic}(e). 
Thus our model exhibits a $T$-independent, incoherent electronic state 
due to the competition between the DE ferromagnetic interaction and 
the SE AF interaction. 

\begin{figure}[t]
 \centerline{\includegraphics[width=7.5truecm]{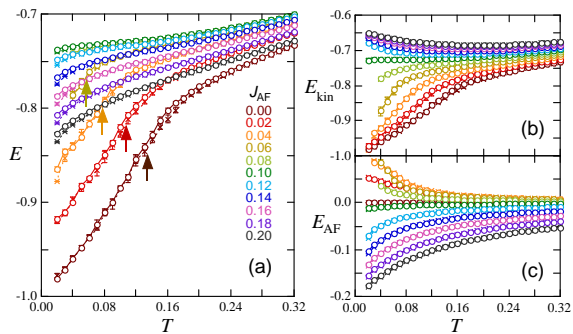}}
\caption{
(Color online)
Temperature dependence of (a) the total internal energy, 
(b) the kinetic energy, and (c) the SE energy. 
Circles and crosses represent the data for $N_{\rm s} = 128$ and $64$, respectively.
In (a), arrows indicate inflection points 
which correspond to the transition to FM state (see text).
}
\label{fig:energy}
\end{figure}

In this critical region, spin fluctuations remain down to low $T$, 
leading to a large amount of residual entropy. 
In Fig.~\ref{fig:energy}(a), we plot the internal energy per site, 
$E = \langle {\cal H} \rangle / N_{\rm s}$, as a function of $T$. 
For small $J_{\rm AF}$, $E$ shows a rapid decrease as $T$ decreases 
and an inflection point (indicated by arrows in the figure),
corresponding to the ferromagnetic transition, 
but the decrease is gradually suppressed as $J_{\rm AF}$ increases. 
At $J_{\rm AF} \sim 0.1$, the $T$ dependence of $E$ becomes minimum, 
whereas it is slowly recovered for larger $J_{\rm AF}$ 
without showing any anomaly associated with a phase transition. 
Hence $E$ shows the smallest $T$ dependence in the critical region. 
Here both the kinetic energy of electrons and the SE energy between localized moments 
become less $T$ dependent, 
as shown in Figs.~\ref{fig:energy}(b) and \ref{fig:energy}(c). 
These results suggest a large amount of entropy remaining at lower $T$, 
since the entropy is given by $T$ integration of the specific heat $dE/dT$ divided by $T$. 
Although the entropy is not well defined in the model including classical vector spins, 
we believe that the almost $T$-independent behavior is robust 
even for more realistic models and that 
the residual entropy is widely observed in the frustrated DE-type models. 
If this PM state survives down to the lowest $T$, 
the system will be a non Fermi liquid 
\cite{Furukawa1994,Biermann2005} 
and show a large mass behavior. 
Another possibility is that some symmetry breaking eventually takes place at a lower $T$ 
beyond our numerical limitation. 

Finally let us discuss the implication of our results to experiments. 
Mo pyrochlore oxides $R_2$Mo$_2$O$_7$ 
are found to 
show puzzling behaviors under external pressure 
in contrast to the $R$-site substitution at ambient pressure, 
that is, 
a collapse of FM and an emergence of the peculiar PM state 
with showing $T$-independent incoherent transport 
\cite{Iguchi2009}. 
These are well reproduced in our results. 
In addition, the experimentally-observed SGM state may be accounted by our PS state: 
Coexisting phases potentially lead to domain formation 
in the presence of long-range interactions such as Coulomb repulsion, and 
the domains pinned by randomness can result in cluster-glass like behaviors
\cite{Dagotto2001}. 
The agreement 
suggests that 
(i) our simple DE model captures an essential physics of the pressure effect 
on the complicated compounds 
and 
(ii) the geometrical frustration plays a major role in the pressure-induced phenomena 
where competition between DE and SE interactions are important. 
These are in sharp contrast to the previous study which focused on 
the relevance of Coulomb repulsion in the $R$-site substitution at ambient pressure 
where the metal-insulator transition is observed
\cite{Solovyev2003}. 
Such observations are important for understanding of 
the distinctive physics between the $R$-site substitution and 
the external pressure. 
Further experiment on the structural change under pressure will help 
to understand the microscopic origin of the difference between the two effects. 
Our result, moreover, predicts 
a non-Fermi-liquid behavior with a large quasiparticle mass or 
some symmetry breaking at low $T$ in the critical region. 
It is highly desired to examine 
experimentally the lower $T$ behavior under pressure in more detail.

%%%%% Summary

To summarize, we have studied the phase competition 
in the double-exchange model on the frustrated pyrochlore lattice. 
In sharp contrast with the unfrustrated models studied for the CMR manganites
which exhibit bicritical behaviors 
between ferromagnetic metal and antiferromagnetic insulator, 
we found that the ferromagnetic metal gradually collapses against 
the super-exchange interaction, and finally a paramagnetic incoherent metal takes over. 
In the critical region, the localized moments are almost uncorrelated down to low temperatures, 
and the electronic transport becomes almost temperature independent and highly incoherent. 
The results are favorably compared with the pressure-induced phenomena in Mo pyrochlore oxides, 
and in addition, predict a large residual entropy. 
These interesting aspects originate from the interplay between 
itinerant electrons and frustrated magnetism. 
Such interplay is widely seen in many frustrated materials and awaits future study.

%%%%% Acknowledgment

We would like to thank Y. Tokura, S. Iguchi, K. Penc, and A. Georges for fruitful discussions. 
This work was supported by Grants-in-Aid for Scientific research 
(Nos. 17071003, 19052008, 17740244, 16GS0219, and 21340090), 
by Global COE Program ``the Physical Sciences Frontier", and 
by the Next Generation Super Computing Project, Nanoscience Program, MEXT, Japan.


\begin{thebibliography}{9}

\bibitem{Zener1951}
C. Zener, Phys. Rev. {\bf 82}, 403 (1951).

\bibitem{Tokura2000}
For a review, {\it ``Colossal Magnetoresistive Oxides"}, edited by Y. Tokura 
(Gordon \& Breach Science Publisher, 2000).

\bibitem{Yunoki1998b}
S. Yunoki and A. Moreo, Phys. Rev. B {\bf 58}, 6403 (1998).

\bibitem{Sen2006}
C. Sen {\it et al.}, Phys. Rev. B {\bf 73}, 224430 (2006).

\bibitem{Yunoki2000}
S. Yunoki, T. Hotta, and E. Dagotto, Phys. Rev. Lett. {\bf 84}, 3714 (2000).

\bibitem{Motome2003}
Y. Motome, N. Furukawa, and N. Nagaosa, Phys. Rev. Lett. {\bf 91}, 167204 (2003).

\bibitem{Diep2005}
For a review, {\it ``Frustrated Spin Systems"}, edited by H. T. Diep 
(World Scientific Publishing, 2005). 

\bibitem{Ali1989}
N. Ali {\it et al.}, J. Solid State Chem. {\bf 83}, 178 (1989).

\bibitem{Katsufuji2000}
T. Katsufuji, H. Y. Hwang, and S-W. Cheong, Phys. Rev. Lett. {\bf 84}, 1998 (2000).

\bibitem{Solovyev2003}
I. V. Solovyev, Phys. Rev. B {\bf 67}, 174406 (2003).

\bibitem{Mirebeau2006}
I. Mirebeau {\it et al.}, Phys. Rev. B {\bf 74}, 174414 (2006).

\bibitem{Iguchi2009}
S. Iguchi {\it et al.}, Phys. Rev. Lett. {\bf 102}, 136407 (2009).

\bibitem{Ikoma2003}
D. Ikoma, H. Tsuchiura, and J. Inoue, Phys. Rev. B {\bf 68}, 014420 (2003).

\bibitem{Anderson1955}
P. W. Anderson and H. Hasegawa, Phys. Rev. {\bf 100}, 675 (1955).

\bibitem{note}
Mo pyrochlore oxides are at quarter filling
when considering the two-fold degeneracy of $e_g^*$ levels 
(one electron per two orbitals). 

\bibitem{Yunoki1998a}
S. Yunoki {\it et al.}, Phys. Rev. Lett. {\bf 80}, 845 (1998).

\bibitem{Poilblanc1991}
D. Poilblanc, Phys. Rev. B {\bf 44}, 9562 (1991).

\bibitem{Gros1992}
C. Gros, Z. Phys. B {\bf 86}, 359 (1992).

\bibitem{Motome2010}
Y. Motome and N. Furukawa, J. Phys.: Conf. Ser. {\bf 200}, 012131 (2010). 

\bibitem{Villain1979}
J. Villain, Z. Phys. B {\bf 33}, 31 (1979). 

\bibitem{Moessner1998}
R. Moessner and J. T. Chalker, Phys. Rev. Lett. {\bf 80}, 2929 (1998).

\bibitem{note2}
Pseudo-gap feature at $\omega \sim 0$ at low $T$ shows the system size dependence; 
the dip becomes small as increasing $N_{\rm s}$.

\bibitem{Furukawa1994}
N. Furukawa, J. Phys. Soc. Jpn. {\bf 63}, 3214 (1994).

\bibitem{Biermann2005}
S. Biermann, L. de' Medici, and A. Georges, Phys. Rev. Lett. {\bf 95}, 206401 (2005).

\bibitem{Dagotto2001}
E. Dagotto, T. Hotta, and A. Moreo, Phys. Rep. {\bf 344}, 1 (2001).

\end{thebibliography}
\end{document}